\documentclass{mem}
\usepackage{natbib}\usepackage{txfonts}\usepackage{balance}
\usepackage{graphicx}
\usepackage[a4paper,breaklinks,dvipdfm]{hyperref}
\idline{75}{282}
\begin{document}
\def\teff{$T\rm_{eff }$}
\def\kms{$\mathrm {km s}^{-1}$}

\title{{\em Gaia} and brown dwarfs from Spain}
\subtitle{}
\author{J.\,A.\,Caballero\inst{1}}
\institute{
Departamento de Astrof\'isica, Centro de Astrobiolog\'ia (CSIC--INTA), 
PO~Box~78, 28691, Villanueva de la Ca\~nada, Madrid, Spain, 
\email{caballero@cab.inta-csic.es}
}

\authorrunning{Caballero}

\titlerunning{{\em Gaia} e nane brune dalla Spagna}

\abstract{
{\em Gaia} will not observe 50\,000 brown dwarfs, but about 100 times less.
However, these $\lesssim$\,500 brown dwarfs will be benchmarks for many substellar topics.
It is possible to identify them in advance and make the list public to all astronomers worldwide through a virtual observatory-compliant ``{\em Gaia} brown dwarf'' catalogue.
This {\em M-, L- and T-dwarf Archive of Interest for Astrophysics} would tabulate precise {\em Gaia} astrometry, multiband photometry, high- and low-resolution spectroscopy and homogeneously derived astrophysical parameters.
Spanish observatories may play a key role in the catalogue preparation.
}
\maketitle{}

\section{Introduction}

The first ``how many brown dwarfs will {\em Gaia} observe'' statement was made before the first brown dwarfs were discovered.
Midway the least massive stars and the most massive exoplanets, the first brown dwarfs were Teide~1 \citep{Re95} and GJ~299\,B \citep{Na95}\footnote{Almost two decades later, there is still some controversy with LP~944--20, GD~165\,B, PPL~15 and some substellar object candidates in Chamaeleon, which had been reported before.}. 
However, a few months before their announcements, during a workshop on future possibilities for astrometry in space, \citet{BB95} proposed finding planets and brown dwarfs with an astrometric interferometric satellite proposed for ESA's Horizon 2000+ programme.
That space mission, originally named {\em GAIA}, was the seed of the current {\em Gaia} observatory.
Since then, there have been several estimations of the number of brown dwarfs that {\em Gaia} will observe.
Based on expected accuracy performances-magnitude relations that were not fixed at that time, \citep{Pe01} estimated ``$\gtrsim$\,50\,000 brown dwarfs'' observable by {\em Gaia}.
Almost 15 years later, this value seems to have spread out throughout innumerable official documents and websites (from ESA internal documents\footnote{\tt http://www.cosmos.esa.int/web/gaia/ science-topics}, through Astrium/Airbus press releases\footnote{\tt http://www.astrium.eads.net/en/news2/ from-hipparcos-to-gaia.html}, to educational brochures for young children\footnote{\tt http://esamultimedia.esa.int/docs/edu/ LittleBooksGaia/English\_everything\_about\_gaia.pdf}).

More realistic estimations of the number of isolated brown dwarfs observable by {\em Gaia} have been available in the literature to the careful reader. 
For example, \citet{HJ02} stated for the first time that {\em Gaia} ``is unlikely to detect field brown dwarfs that have not been already seen in previous near-infrared surveys, to the notable exception of the Galactic plane region''. 
With the advent of new photometric and proper-motion surveys at low Galactic latitudes, performed with e.g. VISTA or {\em WISE} data, the room for {\em Gaia} new discoveries of brown dwarfs gets tiny.  
Furthermore, \citet{Sm08} claimed that ``{\em Gaia} will only observe 25\,\% of L and T dwarfs within 50\,pc, which [...] amounts to less than 400 objects''.

Up to now, the most precise estimation of the number of isolated brown dwarfs observable by {\em Gaia} is that of \citet{Sa13}.
Using the faintest magnitude limit at $G$ = 20\,mag, the spatial densities of field ultracool dwarfs of \citet{Ca08} and theoretical spectral energy distributions for known spectral types, they estimated that {\em Gaia} will detect around 400\,000 M5--L0, 600 L0--5, 30 L5--T0 and 10 T0--8 dwarfs.
New re-estimations based on SDSS photometry and the DwarfArchive\footnote{\tt http://dwarfarchive.org} suggest that these values are approximately correct, which supports the \citet{HJ02} and \citet{Sm08}'s original calculations (\citealt{Sa14}; \citealt{Sm14}).
However, there are very few young field M dwarfs with masses below the hydrogen burning limit and, therefore, that can be classified as brown dwarfs (mostly members in young moving groups), while most of the field early L dwarfs are actually very low-mass stars above the substellar boundary \citep{Ki05}.
Because of it, {\em Gaia} will be able to actually observe no more than $\sim$500 isolated brown dwarfs, two orders of magnitudes less than originally considered (unless the {\em Gaia} limit magnitude is extended to $G$ = 21\,mag, for which the satellite would detect about 2\,000 brown dwarfs).

\section{MAIA}

\subsection{The catalogue}

In spite of the apparently scarce number of ``{\em Gaia} brown dwarfs'', because of their brightness, the $\lesssim$\,500 of them will be benchmarks for many topics in substellar astronomy, such as atmosphere models, field mass luminosity/function, multiplicity, moving groups, ages, metallicity (in multiple systems), formation and dynamical evolution.
Identifying and cataloguing these $\lesssim$\,500 brown dwarfs in advance of the {\em Gaia} data releases seemed to be an obvious thought before the workshop, since it was not only proposed in this contribution, but also in another one by \citet{Mo14}.  

This catalogue should (or shall) be public, as useful as possible for astronomers worldwide, and virtual-observatory compliant.
It should collect homogeneous low- and high-resolution spectroscopy and multi-band photometry obtained with ground telescopes, together with {\em Gaia} parallaxes and proper motions, from which deriving accurate luminosities, effective temperatures, surface gravities, radii, spatial densities and Galactocentric velocities, among other key parameters.
In a sense, it would resemble a mixture of the Gaia-ESO survey archive \citep{Gi12, Ra13} and the exoplanet encyclopaedia\footnote{\tt http://exoplanet.eu} \citep{Sc11}.

There have already been some attempts of starting the construction of a ``{\em Gaia} brown dwarf'' catalogue, but at a much smaller scale.
One of them is publicly available at the conference website\footnote{\tt http://gaiabds.oato.inaf.it} and has hyperlinks to spectroscopic observations with Keck and ESO telescopes (c.f., \citealt{Sm14}).  

Creating a ``{\em Gaia} brown dwarf'' catalogue is not a one-man's task: it requires the close collaboration between experts in data archiving and mining, low- and high-resolution spectroscopy, multi-band photometry and spectral energy distributions, astrophysical properties of ultracool dwarfs, and {\em Gaia} data analysis.
Moreover, given the 750\,MEUR total cost of the space mission, fully paid by ESA (including manufacture, launch and ground operations, but not counting the manpower of astronomers), and the relevance of ultracool stellar and substellar studies in the 21st century Astronomy, one may wonder if establishing a small team of European astronomers devoted to create a ``{\em Gaia} brown dwarf'' catalogue is not a moral obligation. 
I suggest the name MAIA, from {\em `M-, L- and T dwarf Archive of Interest for Astrophysics'}, for the catalogue.
As seen below, MAIA could easily be expanded and contain not only isolated field brown dwarfs in the solar neighbourhood.

\subsection{The targets}

The MAIA primary targets would be the few dozen T and L5--9 dwarfs that {\em Gaia} will be able to observe (such as $\epsilon$~Ind\,Bab, Luhman~16\,B, 2M0758+32 or 2M1254--01), some of the brightest 200--300 L0--5 field dwarfs and a few selected nearby young M dwarfs in moving groups (e.g., LP~944--20 M9\,V and other high-mass brown dwarfs in AB~Doradus, TW~Hydrae, Tucana-Horologium).

The selection criterion must be as homogeneous as possible.
The simplest way may be just cataloguing all {\em Gaia} L and T dwarfs plus the few already-known young M-type brown-dwarf candidates in the first data release (22 months after launch); MAIA would thus contain both brown dwarfs and very low-mass stars. 
Another option is to apply a spectral type-dependent $G$-magnitude-limit cut (which roughly translates into a distance-limit cut). 
Such a ``staired'' selection criterion is already in use for the CARMENES\footnote{CARMENES is a HARPS-like spectrograph specifically designed for detecting low-mass exoplanets around M dwarfs \citep{Qu12}} input catalogue (i.e., M0\,V $J <$ 7.0\,mag, M1\,V $J <$ 7.5\,mag, M2\,V $J <$ 8.0\,mag, etc. -- \citealt{Ca13}) and warrants  that the expensive time of any high-resolution spectroscopic study is actually devoted to the brightest stars/brown dwarfs of each spectral type.

\subsection{The data}

For each brown dwarf, MAIA should tabulate the following data:
\begin{itemize}
\item Identifier and discovery name,
\item {\em Gaia} coordinates and proper motions (and parallactic distances after second release),
\item radial velocities (from high-resolution spectroscopy with ground telescopes),
\item Galactocentric spatial velocities (and potential membership in kinematic group),
\item magnitudes in as much photometric systems as possible (in general from SDSS $ugriz$, through {\em Gaia} BP and RP and 2MASS $JHK_{\rm s}$, to {\em WISE} $W$1--4),
\item wide multiplicity data (if the brown dwarf is a common proper-motion companion to a brighter star, one can extrapolate certain parameters for the brown dwarf -- e.g., metallicity, more precise distance),
\item close multiplicity data (roughly one out of five field brown dwarfs is a close binary),
\item activity indicators (pseudo-equivalent widths of H$\alpha$ and calcium triplet emission, X-rays),
\item rotational velocity,
\item photometric period,
\item hyperlinks to public low- and high-resolution spectroscopic data,
\item if possible, hyperlinks to public high-resolution imaging data,
\item all homogeneously derived astrophysical parameters ($T_{\rm eff}$, $\log{g}$, $R$, $L$, some reliable [Fe/H] proxy, age, etc. -- or, at least, as homogeneous as possible if the origin of the spectrophotometric data is heterogeneous),
\item any remark relevant for the reader,
\item and references for each item.
\end{itemize}
Photometry and spectral energy distribution fitting provides a first-order estimation of effective temperature.
Low-resolution spectroscopy provides a better $T_{\rm eff}$ estimation, as well as hints to gravity, activity and metallicity, which in turn give information on, e.g., age, multiplicity and population membership.
However, the data that best complement the accurate {\em Gaia} parallaxes and proper motions are the high-resolution spectroscopic ones.

For measuring precise radial and rotational velocities, reliable metallicities and abundances and, especially, effective temperatures and surface gravities in ultracool dwarfs, European astronomers will soon be able to use a suite of near-infrared spectrographs with large wavelength coverage and moderate and high resolutions:
X-Shooter \& CRIRES+/8.2\,m~VLT,
PHOENIX/4\,m~Mayall,
GIANO/3.6\,m~TNG,
ESPaDOnS \& SPIRou/3.6\,m~CFHT,
CAR\-ME\-NES/3.5\,m~Calar Alto,
and NET/2.6\,m~NOT.
Not by chance, most of the near-infrared spectrographs in the Northern Hemisphere are located {\em in Spain}, which partly explains the title of this contribution.

\subsection{Potential upgrades} 

The other part of explanation of the Spanish title is the {\em Red espa\~nola de Explotaci\'on de Gaia}\footnote{\tt https://gaia.am.ub.es/Twiki/bin/view/
RecGaia/WebHome}, which is a network of researchers from virtually all astronomy centres in Spain.
One of the research lines of the network is ``very low-mass stars, brown dwarfs and exoplanets'', which aims at coordinating (or at least avoiding overlapping of) the efforts of researchers in Spanish institutions on ultracool dwarfs.
Of relevance for this contribution, there are teams working on astrometry of known radial-velocity systems with brown dwarfs, radial velocity of new astrometric systems with brown dwarfs, L and T dwarfs, both isolated or as companions, and young brown dwarfs at the bottom of the (initial) mass function in young open clusters and stellar associations.

MAIA may actually be a collection of ``{\em Gaia} brown dwarf'' catalogues.
Some possible additions to the field brown dwarfs may include:
\begin{itemize}
\item {\em All L-type objects} observed by {\em Gaia}, regardless they are stars or brown dwarfs.
Compiling high-resolution spectroscopy of each target may not be feasible. 
\item {\em Brown dwarf companions to stars.}
It is possible to investigate in detail faint brown-dwarf proper-motion companions to bright stars.
Although the companions will not be detected by {\em Gaia}, the primaries will be bright enough to have accurate {\em Gaia} parallaxes and proper motions and, perhaps, optical high-resolution spectroscopy.
\item {\em Young brown dwarfs in open clusters}.
Sadly, {\em Gaia} will be able to observe in average a dozen brown dwarfs per cluster.
Key parameters to take into account are age, distance, surface density, extinction and background, which make the Pleiades, Upper Scorpius and $\sigma$/$\lambda$~Orionis probably the best regions to look for young brown dwarfs with {\em Gaia}.
The real {\em Gaia} input for young brown dwarfs will be the determination of precise heliocentric distances to clusters, from which deriving accurate mass functions together with photometry and the most adavnced theoretical models.
\item {\em M dwarfs for exoplanet searches}. 
Lastly, MAIA can be extended by including the brightest M dwarfs in both hemispheres suitable for radial-velocity monitoring with high-resolution spectrographs (HPF, CARMENES and SPIRou in the North, CRIRES+ and perhaps a new instrument at the NTT in the South).
\end{itemize}

\begin{acknowledgements}
Financial support was provided by the Spanish Ministerio de Ciencia e Innovaci\'on under grant AYA2011-30147-C03-03.	
\end{acknowledgements}

\bibliographystyle{aa}

\end{document}